\documentclass[aps,prl,superscriptaddress,twocolumn,twoside,a4paper,floatfix]{revtex4}

\usepackage{hyperref,graphicx,amsmath,latexsym,amssymb,verbatim,revsymb,color,mathpazo}
\newcommand{\ra}{\rangle} 
\newcommand{\la}{\langle} 
\newcommand{\Tr}{\textrm{Tr}} 
\newcommand{\bra}[1]{\langle #1|} 
\newcommand{\ket}[1]{|#1\rangle} 
\newcommand{\Exp}[1]{\langle#1\rangle}

\begin{document}

\title{The smallest possible heat engines}

\author{Noah Linden} \affiliation{Department of Mathematics, University of Bristol$\text{,}$ University Walk, Bristol BS8 1TW, United Kingdom} 
\author{Sandu Popescu} \affiliation{H. H. Wills Physics Laboratory, University of Bristol$\text{,}$ Tyndall Avenue, Bristol, BS8 1TL, United Kingdom} 
\author{Paul Skrzypczyk} \affiliation{H. H. Wills Physics Laboratory, University of Bristol$\text{,}$ Tyndall Avenue, Bristol, BS8 1TL, United Kingdom}

\date{\today}
\begin{abstract}
We construct the smallest possible self contained heat engines; one composed of only two qubits, the other of only a single qutrit. The engines are self-contained as they do not require external sources of work and/or control. They are able to produce work which is used to continuously lift a weight. Despite the dimension of the engine being small, it is still able to operate at the Carnot efficiency.
\end{abstract}

\maketitle

Very recently considerable progress has been made in understanding foundational aspects of thermodynamics by addressing a new class of questions: whether there exist additional fundamental limitations on thermal machines, arising specifically due to their size \cite{lps}; whether there is an absolute minimum size to all thermal machines; alternatively, if machines are small, is their performance constrained,  for example can they still achieve the Carnot efficiency?

For the case of refrigerators it was shown \cite{lps} that there is no minimum size when self-contained machines consisting of only two qubits, or a single qutrit, were discovered. Furthermore it was shown that these refrigerators have two important properties; they are able to cool towards absolute zero \cite{lps} and they can operate at the Carnot efficiency \cite{sblp}, showing that here size appears not to limit the machines at all.

One advantage in focusing initially on refrigerators rather than heat engines is because they allowed us to avoid the explicit notion of {\em work} in our study, whilst making progress in understanding small thermal machines. Work is a concept that appears difficult to capture in the quantum regime, where ideas such as ``order'' usually associated to it are hard to make sense of. Therefore it is not clear that the existence of small self-contained refrigerators implies anything about small self-contained heat engines. 

The study of the thermodynamics of quantum systems has become a thriving and dynamic field \cite{QT1,QT2,QT3}. The definition of work has indeed been discussed extensively \cite{W1,W2,W3} and applied to quantum heat engines \cite{H1,H2,H3,H4,H5,H6,H7} and Carnot cycles \cite{C1,C2,C3,C4,C5,C6}. The focus however has been on quantum thermal machines which explicitly or implicitly have macroscopic objects in the background which supply either work or some form of control; for example systems which are externally driven, or make use of sequences of unitary evolutions. The definitions of work given so far have been applicable and very well suited to this situation. Our focus is however different; we are interested here in \emph{self contained} heat engines, where there is no external work or control -- the only external interaction being with thermal reservoirs. The previous definitions of work therefore do not apply directly to our situation, and an alternative must be found.

In this work we will use exactly the definition of work put forward by Carnot, namely: 

\begin{quote}
	{\it ``motive power (work) is the useful effect that a motor is capable of producing. This effect can always be likened to the elevation of a weight to a certain height.'' }\cite{carnot}
	\end{quote}
to construct a quantum heat engine. Our engine will produce work in the same way as above, by lifting a weight; here the weight consists of a system with an infinite number of energy levels, and the work created by our heat engine causes the position of the weight to increase with time. We will present two models for the heat engine, one consisting of two qubits (two level systems), and a second consisting of a single qutrit (three level system). We will also show that they can operate at the Carnot efficiency.

\section{Two-Qubit Model}
The heat engine consists of two qubits. Qubit 1 is in contact with a ``cold" heat bath at temperature $T_c$ and qubit 2 is in contact with a ``hot'' bath at temperature $T_h$.  Like every heat engine, this engine acts by extracting heat from the hot bath and converting it into work, while dumping some heat into the cold bath.

We imagine that the engine delivers work by pulling up a weight. The weight is isolated from both baths. To simplify the situation we consider that the weight is pulled up very slowly, so that we can neglect the change of its kinetic energy and consider only the potential energy. Furthermore, we suppose that the weight can be situated only at some discrete equidistant heights, so that the energy difference between them is the same. Hence, the weight has discrete energy eigenstates $|n\ra_w$ with corresponding energy eigenvalues $E_n=n\cal E$, with ${\cal E}>0$. Alternatively, we can imagine that the engine delivers work by pumping energy into a harmonic oscillator; both situations are formally almost equivalent (the harmonic oscillator energies are limited from below by 0).

\begin{figure}[h] 
	\includegraphics[width=0.75\columnwidth]{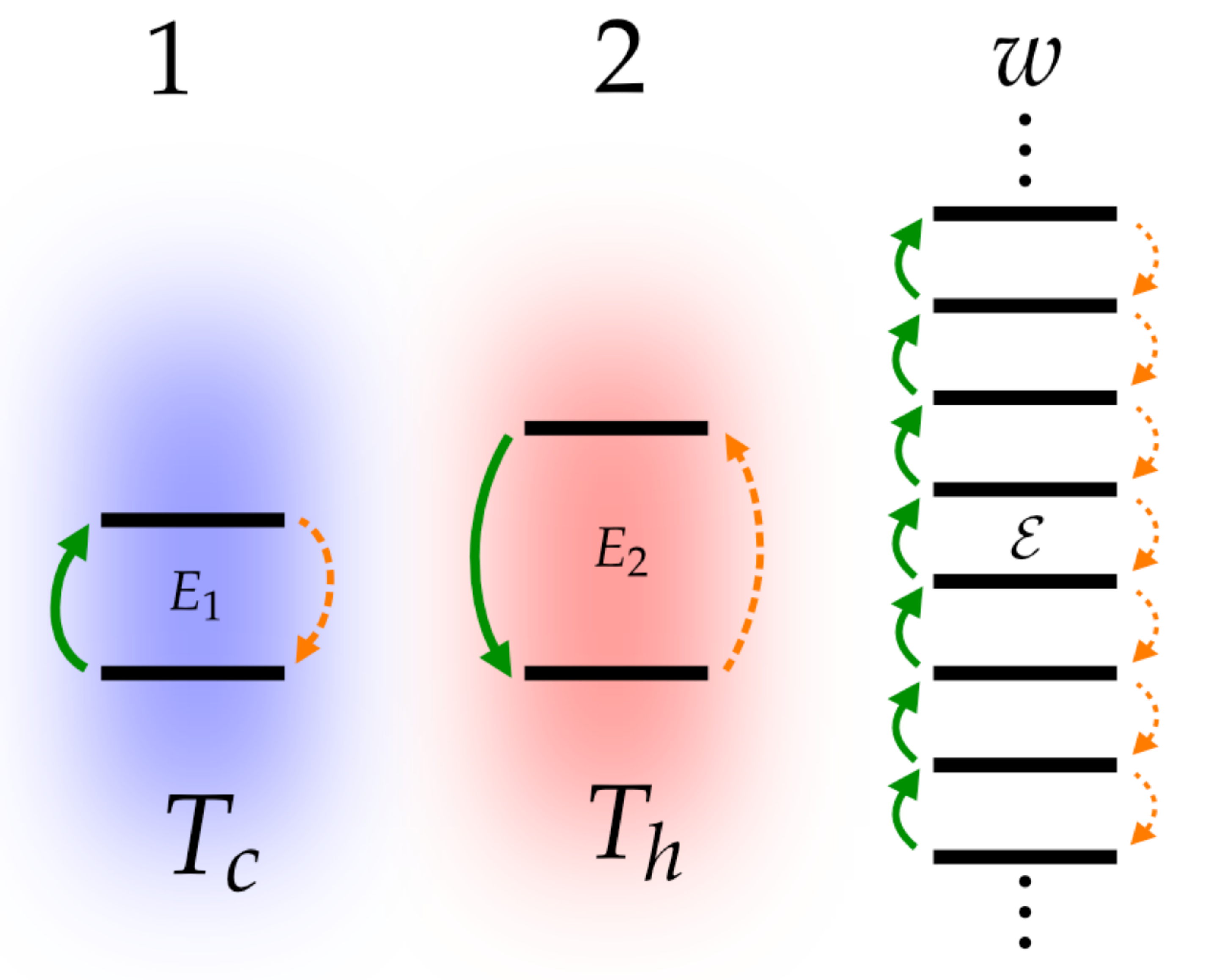} \caption{\label{f:schematic qubit} {\bf Schematic diagram of two-qubit heat engine.} Qubits 1 and 2, with energy level separations $E_1$ and $E_2$ are in contact with thermal reservoirs at temperatures $T_c$ and $T_h$ respectively. To this a weight is connected, with separation $\mathcal{E} = E_2 - E_1$. The particles interact with each other and the temperatures are chosen such that the transition where the weight is lifted (solid green arrows) is biased over the transition where the weight falls (dashed orange arrows). } 
\end{figure}

In the absence of interaction we take qubit $i$, $i=1,2$  to have energy eigenstates $|0\ra_i$ and $|1\ra_i$ and corresponding energy eigenvalues $0$ and $E_i>0$. The free Hamiltonian of the system is thus
\begin{equation}\label{e:free Hamiltonian}
	H_0=E_1\ket{1}_1\bra{1}+E_2\ket{1}_2\bra{1} + \sum_{n=-\infty}^{\infty} n{\cal E}\ket{n}_w\bra{n}
\end{equation}
The energies are taken such that
\begin{equation} 
	E_2 - E_1 = \mathcal{E}.\label{energy_conservation}
\end{equation}
Given this constraint, the energy levels $|01,n\ra$ and $|10,n+1\ra$ are degenerate. The engine acts by making transitions between these degenerate states. 

The qubits and weight interact via the Hamiltonian
\begin{equation} 
	H_{int}= g\!\!\!\!\sum_{n=-\infty}^{\infty}\!\!\Big(|01,n\ra\la 10, n+1| + |10,n+1\ra\la 01, n|\Big). 
\end{equation}

The basic idea behind the design of this engine is to bias the transition $|01,n\ra \rightarrow|10,n+1\ra$  in which the weight is lifted in favor of the reverse transition  in which the weight is lowered. This is obtained by coupling the two qubits to heat baths at different temperatures, $T_h> T_c$, chosen such that the probability for the qubits to be initially in the state $|01\ra$ is larger than the probability to be in the state $|10\ra$. 
 
In the absence of interaction each of the two qubits is at thermal equilibrium. Let the probabilities for the ground and excited state of qubit $i$ be $r_i$ and $\overline{r_i}$. At thermal equilibrium they are related by 
\begin{align}\label{e:thermal def}
	\overline{r_1}&=r_1 e^{-{{E_1}\over{kT_c}}},& \overline{r_2}&=r_2 e^{-{{E_2}\over{kT_h}}},
\end{align}   
where $k$ is Boltzmann's constant. This leads to the probabilities $q_{10}$ of the state $\ket{10}$ and $q_{01}$ of $\ket{01}$ to be
\begin{align} 
	q_{10}&=\overline{r_1}r_2=r_1r_2e^{-{{E_1}\over{kT_c}}},& q_{01}&= r_1\overline{r_2}=r_1r_2 e^{-{{E_2}\over{kT_h}}}.
\end{align} 
Therefore in the absence of interaction, the probability of the state  $|01\ra$ is larger than that of $|10\ra$, i.e.
 $q_{01}>q_{10}$, when we have
\begin{equation}
	\frac{E_1}{T_c}> \frac{E_2}{T_h}.\label{e:bias}
\end{equation}

To demonstrate the functioning of our engine we also need to model the interaction of the two qubits with their heat baths. We take here a simple model, the same as the one used in \cite{lps}. Namely, we consider that in a short time interval $\delta t$ with probability $p_i\delta t$ the state of qubit $i$ is reset to its thermal state, the density matrix $\tau_i$,
\begin{equation} \label{e:tau}
	\tau_i= r_i|0\ra_i\la 0|+\overline{r_i}|1\ra_i\la 1|.
\end{equation}
The overall equation of motion is 
\begin{equation}
    \label{eqn-of-motion1} \frac{
    \partial \rho}{
    \partial t} = -i[H_0+H_{int}, \rho] + \sum_{i=1}^2 p_i(\tau_i {\rm Tr}_i \rho - \rho)
\end{equation}
where $\rho$ denotes the state (density matrix) of the full system comprised of qubits 1 and 2 and the weight. This Master equation is consistent in the limit where $g$ and $p_i$ are small, otherwise there are corrections to the dissipative dynamics which must be take into account \cite{T1}. In this regime we expect our device will act as a heat engine and continuously lift the weight as long as the relations \eqref{energy_conservation} and \eqref{e:bias} are fulfilled.

The quantity of interest is the average energy of the weight, $\langle E_w\rangle$. We expect that, after a transient period, this average energy will increase uniformly, that is we expect that
\begin{equation} \frac{d}{dt}\Exp{E_w}=\text{const} >0
\end{equation}
We now show that this is indeed the case. From \eqref{eqn-of-motion1} we obtain
\begin{equation}\label{e:ExpEw}
\frac{d}{dt}\Exp{E_w} = \frac{d}{dt} \Tr(H_w\rho)
= -ig\mathcal{E}\Delta(t)
\end{equation}
where 
\begin{equation}
	\Delta(t)\! = \!\sum_n\!\Big(\!\bra{01,n}\rho\ket{10,n+1}\! - \!\bra{10,n+1}\rho\ket{01,n}\!\Big)
\end{equation}
If we define further the two quantities $\Gamma_1(t)$ and $\Gamma_2(t)$ which are the instantaneous ground state probabilities for qubits 1 and 2 respectively,
\begin{eqnarray}
	 \Gamma_1(t) &=& \sum_n\Big(\bra{00,n}\rho\ket{00,n} + \bra{01,n}\rho\ket{01,n}\big) \\
	\Gamma_2(t) &=& \sum_n\Big(\bra{00,n}\rho\ket{00,n} + \bra{10,n}\rho\ket{10,n}\big)
\end{eqnarray}
then together, along with $\Delta(t)$, these three quantities obey the coupled set of equations
\begin{eqnarray}
	\frac{d}{dt}\Delta(t) &=& 2ig\big(\Gamma_1(t)-\Gamma_2(t)\big) - (p_1+p_2)\Delta(t) \nonumber \\
	\frac{d}{dt} \Gamma_1(t) &=& +ig\Delta(t) + p_1\big(r_1 - \Gamma_1(t)\big) \\
	\frac{d}{dt}\Gamma_2(t) &=& -ig\Delta(t) + p_2\big(r_2 - \Gamma_2(t)\big) \nonumber
\end{eqnarray}
This set of equations can easily be solved. The solution for $\Delta(t)$ is given by
\begin{equation}\label{e:Delta sol}
	\Delta(t) = \sum_{i=1}^3\delta_i e^{\lambda_i t} + \frac{2igp_1p_2(r_1-r_2)}{(p_1+p_2)(2g^2 + p_1p_2)}
\end{equation} 
where the $\delta_i$ are constants determined by the initial conditions. The eigenvalues $\lambda_i$ are the solutions of the characteristic equation $(\lambda+p_1)(\lambda+p_2)(\lambda+p_1+p_2) + 2g^2(2\lambda + p_1+p_2) = 0$, and all have negative real part. The first term in \eqref{e:Delta sol} is therefore the transient behaviour which decays in time, and thus in the long time limit only the second term survives and $\Delta$ becomes a constant. Thus, from equation \eqref{e:ExpEw},
\begin{equation}
	\lim_{t\to \infty} \frac{d}{dt}\Exp{E_w} = \frac{2\mathcal{E}g^2p_1p_2(r_1-r_2)}{(p_1+p_2)(2g^2 + p_1p_2)}
\end{equation}
which is positive whenever $r_1 > r_2$, which is equivalent to $q_{10} > q_{01}$ -- the condition \eqref{e:bias} imposed initially to ensure  a probability bias in the correct direction. Therefore the average energy of the weight increases linearly with time, thus demonstrating that the weight is able to be continuously lifted. 

We note however that the probability distribution of the position of the weight, (and therefore the probability distribution of its energy) will be expected to also spread in time. Essentially we expect the evolution, after the transient period, to be that of a biased random walk. This can be confirmed to be the actual behaviour of the system. A similar (but more involved) calculation to the above shows that $\Delta E_w^2$ goes like $t$ in the limit that $t$ tends to infinity. Thus the standard deviation in the energy, $\Delta E_w$, varies asymptotically as $\sqrt{t}$, as expected for a random walk.
\section{The Efficiency}
The most important property of the engine is the efficiency with which it can run, namely the amount of work that can be delivered for a given amount of heat extracted from the hot bath. 

The rate at which heat flows between the qubits and their environments is given by the change in energy of each qubit due to the interaction with the baths. Given the Master equation \eqref{eqn-of-motion1}, the heat currents are therefore
\begin{equation}\label{e:heat}
	\frac{d}{dt}Q_i = p_i\Tr\Big(H_i\big(\tau_i-\rho_i(t)\big)\Big)
\end{equation}
Recall that $\Gamma_1(t)$ and $\Gamma_2(t)$ are the ground state populations of $\rho_1(t)$ and $\rho_2(t)$ respectively. The asymptotic solution for these two variables is given by
\begin{equation}\label{e:gamma sol}
	\lim_{t\to\infty} \Gamma_i(t) = \frac{p_1p_2(p_1+p_2) r_i + 2g^2(p_1r_1+p_2r_2)}{(p_1+p_2)(2g^2+p_1p_2)}
\end{equation}
Substitution of this result into \eqref{e:heat}, and making use of \eqref{e:free Hamiltonian} and \eqref{e:tau} it is found that
\begin{equation}
	\frac{d}{dt}Q_i = (-1)^i \frac{2E_ig^2p_1p_2(r_1-r_2)}{(p_1+p_2)(2g^2 + p_1p_2)}
\end{equation}
indicating, as expected, that heat is flowing out of qubit 1 into the `cold' reservoir and is flowing into qubit 2 from the `hot' reservoir.
Identifying now $\tfrac{d}{dt}\Exp{E_w}$ as the work extracted from the engine, since this is the energy which is used to raise the weight, it is seen that the efficiency $\eta^Q$ of the quantum heat engine is given by
\begin{equation}
	\eta^Q = \frac{\tfrac{d}{dt}\Exp{E_w}}{\tfrac{d}{dt}Q_2} = \frac{E_2-E_1}{E_2} = 1-\frac{E_1}{E_2}
\end{equation}
However, for the weight to be lifted, condition \eqref{e:bias} must be satisfied. Thus, in the limit where the speed at which the weight is lifted becomes infinitely slow, that is when \eqref{e:bias} becomes an equality, the quantum fridge reaches its maximum possible efficiency $\eta^Q_{\text{max}}$, given by
\begin{equation}
	\eta^Q_{\text{max}} = 1-\frac{T_c}{T_h}
\end{equation}
which is exactly the Carnot efficiency of a heat engine running between two reservoirs at temperatures $T_c$ and $T_h$. Thus no new additional constraints arise due to the small size of the engine presented, with the only constraint upon its efficiency being the Carnot limit. 
\section{One-Qutrit Model}

\begin{figure}[t] 
	\includegraphics[width=0.75\columnwidth]{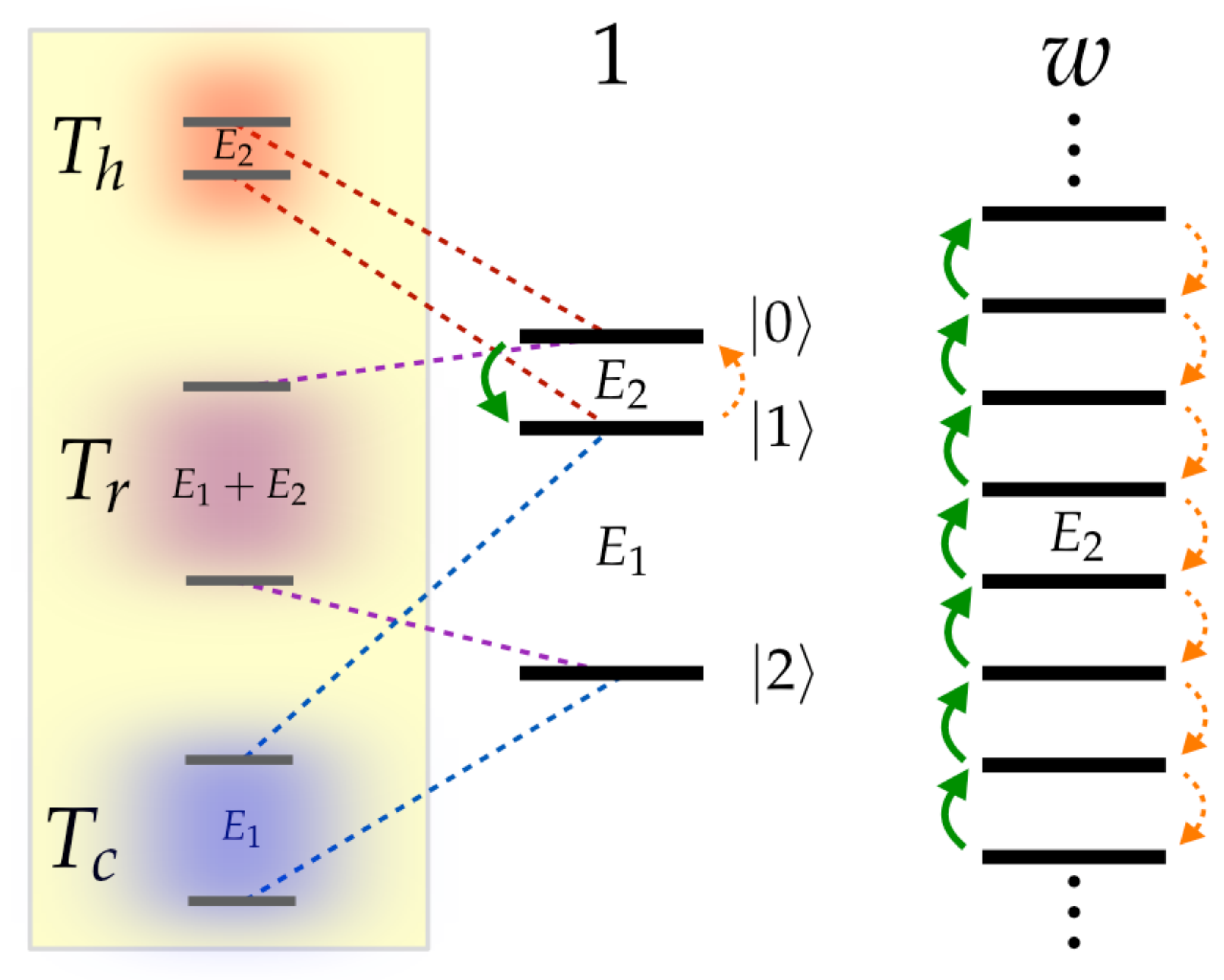} \caption{\label{f:schematic qutrit} {\bf Schematic diagram of qutrit heat engine.} The qutrit is assumed to have each of its transitions at different temperatures. The temperature of the transitions is shown in the inset on the left. A weight is again connected, with energy level separation $E_2$, matching that of the separation of the two upper levels of the qutrit. The particles interact with each other and the temperatures are chosen such that the transition where the weight is lifted (solid green arrows) is biased over the transition where the weight falls (dashed orange arrows). } 
\end{figure}
There is a yet smaller self contained heat engine than the one described above. It consists of a single qutrit -- a quantum system with a three dimensional Hilbert space. Imagine that the three transitions  (from $|0\ra$ to $|1\ra$, from $\ket{0}$ to $\ket{2}$, and from $\ket{1}$ to $\ket{2}$) are affected by different heat baths. This might occur, for example, due to the physical geometry of the system comprising the qutrit. Given such a set-up, it is possible to construct a smaller heat engine. The overall set-up can be seen in the Schematic diagram in Fig.~\ref{f:schematic qutrit}.

The corresponding Master equation for this system can also be written down and solved for the long-time behaviour of the system. The detailed calculation can be found in the Appendix. The solution, in the situation where all thermalisation rates are equal, denoted here by $p$, is found to be
\begin{multline}
\lim_{t\to\infty}\frac{d}{dt}\Exp{E_w} = \\ \frac{4g^2E_2p(\overline{r_r}\,\overline{r_h}-\overline{r_c}r_h)}{2g^2(2+\overline{r_c}+\overline{r_r})+3p(3-\overline{r_c}\,\overline{r_r} - r_c\overline{r_h}-r_rr_h)}
\end{multline}
where $r_c$, $r_r$ and $r_h$ are the ground state thermal populations of qubits at the corresponding temperatures, defined as in \eqref{e:thermal def}, and $g$ is again the interaction strength. The condition that the average energy of the weight asymptotically increases in time is now given by 
\begin{equation}
	\overline{r_r}\,\overline{r_h} > \overline{r_c}r_h
\end{equation}
which says that, in the thermal states, the product of probabilities for state $\ket{2}$ must exceed that of the product of  probabilities of state $\ket{1}$.

{\bf Acknowledgements} We acknowledge support from EU integrated project Q-ESSENCE.

\section{Appendix: One-Qutrit Model Details}
In the absence of interaction, the free Hamiltonian of the qutrit and weight is given by
\begin{equation}
	H_0 = E_1\ket{1}_1\bra{1} + (E_1+E_2)\ket{2}_1\bra{2} + E_2\sum_n n\ket{n}_w\bra{n}
\end{equation}
The qutrit and weight interact with each other via the interaction Hamiltonian
\begin{equation}
	H_{int} = g\sum_n \Big(\ket{1,n+1}\bra{2,n} + \ket{2,n}\bra{1,n+1}\Big).
\end{equation}
The qutrit, due to its physical geometry, has each of its transitions in contact with a different thermal reservoir. The transition between the states $\ket{0}$ and $\ket{1}$ is at temperature $T_c$, the transition between the states $\ket{0}$ and $\ket{2}$ is at the temperature $T_r$ and the transition between the states $\ket{1}$ and $\ket{2}$ is at the temperature $T_h$. The Master equation governing the dynamics of the qutrit interacting with the environments and weight is given by
\begin{eqnarray}\label{eqn-of-motion2} 
	\frac{\partial \rho}{\partial t} &=& -i[H_0+H_{int}, \rho]  \\
	&+&  p_c\Big(\tau_c\bigl(\bra{0}\rho\ket{0}+\bra{1}\rho\ket{1}\big) + \ket{2}\bra{2}\bra{2}\rho\ket{2} - \rho\Big)\nonumber \\
	&+&  p_r\Big(\tau_r\big(\bra{0}\rho\ket{0}+\bra{2}\rho\ket{2}\big) + \ket{1}\bra{1}\bra{1}\rho\ket{1} - \rho\Big)\nonumber \\
	&+&  p_h\Big(\tau_h\bigl(\bra{1}\rho\ket{1}+\bra{2}\rho\ket{2}\big) + \ket{0}\bra{0}\bra{0}\rho\ket{0} - \rho\Big)\nonumber
\end{eqnarray}
where 
\begin{eqnarray}
\tau_c = r_c\ket{0}\bra{0} + \overline{r_c}\ket{1}\bra{1}, && \tau_r = r_r\ket{0}\bra{0}+\overline{r_r}\ket{2}\bra{2},\nonumber \\
	\tau_h = r_h\ket{1}\bra{1} &+& \overline{r_h}\ket{2}\bra{2} \nonumber
\end{eqnarray}
are the thermal states for each of the three transitions respectively, where the $r_i$ are defined according to \eqref{e:thermal def}.

To show that the qutrit engine is also able to produce work, the quantity of interest again is the asymptotic behaviour of the mean energy of the weight. From \eqref{eqn-of-motion2} it follows that
\begin{equation}
	\frac{d}{dt}\Exp{E_w} = \frac{d}{dt}\Tr(H_w\rho) = -igE_2\Omega(t)
\end{equation}
where
\begin{equation}
	\Omega(t) = \sum_n \Big( \bra{2,n}\rho\ket{1,n+1} - \bra{1,n+1}\rho\ket{2,n}\Big)
\end{equation}
Similarly to the two-qubit case, by defining further the two variables $B_1(t)$ and $B_2(t)$, the instantaneous excited state probabilities of the qutrit,
\begin{align}
	B_1(t) &= \sum_n\bra{1,n}\rho\ket{1,n},& B_2(t) &= \sum_n\bra{2,n}\rho\ket{2,n}
\end{align}
then together with $\Omega(t)$ they satisfy the following set of equations
\begin{eqnarray}
\frac{d}{dt}\Omega &=&+2ig(B_2-B_1) - (p_c+p_r+p_h)\Omega  \\
%\!\!\!\tfrac{d}{dt}B_1\! &=& \!-ig\Omega - (p_c+p_h\overline{r_h})B_1 - (p_c\overline{r_c}-p_hr_h)B_2 + p_c\overline{r_c} \nonumber \\
\frac{d}{dt}B_1 &=& -ig\Omega +p_c(\overline{r_c}(1-B_2)-B_1) - p_h(\overline{r_h}B_1-r_hB_2) \nonumber \\
%\!\!\!\tfrac{d}{dt}B_2\! &=& \!+ig\Omega - (p_r\overline{r_r}-p_h\overline{r_h})B_1 -(p_r + p_hr_h)B_2 + p_r\overline{r_r}\nonumber \\ 
\frac{d}{dt}B_2 &=& +ig\Omega +p_r(\overline{r_r}(1-B_1) - B_2) + p_h(\overline{r_h}B_1 - r_hB_2)\nonumber
\end{eqnarray}
The behaviour is found to be qualitatively the same as the two-qubit heat engine; after an initial transient period, $\Omega(t)$ tends to a constant. Thus the long-time behaviour of the mean energy of the weight is given by 
\begin{widetext}
\begin{equation}
\lim_{t\to\infty}\frac{d}{dt}\Exp{E_w} = 	\frac{2g^2E_2(p_cp_r(\overline{r_r}-\overline{r_c})-p_h(r_h-\overline{r_h})(p_c\overline{r_c}+p_r\overline{r_r}))}{2g^2(p_c(1+\overline{r_c}) + p_r(1+\overline{r_r}))+(p_c+p_r+p_h)(p_cp_r(1-\overline{r_c}\,\overline{r_r})+p_cp_h(1-r_c\overline{r_h})+p_rp_h(1-r_rr_h))}
\end{equation}
\end{widetext}
and thus the weight is lifted whenever the following condition is satisfied
\begin{equation}
	p_cp_r(\overline{r_r}-\overline{r_c})>p_h(r_h-\overline{r_h})(p_c\overline{r_c}+p_r\overline{r_r}).
\end{equation}


\begin{thebibliography}{99} 
	\bibitem{lps} N. Linden, S. Popescu and P. Skrzypczyk, Phys. Rev. Lett. {\bf 105}, 130401 (2010).
	\bibitem{sblp} P. Skrzypczyk, N. Brunner, N. Linden and S. Popescu, arXiv:quant-ph/1009.0865
	\bibitem{carnot} S.~Carnot, {\it R\'{e}flexions sur la puissance motrice du feu} (1824); {\it Reflection on the Motive Power of Fire}, Dover (1960).
	\bibitem{QT1} G.~Gemma, M.~Michel and G.~Mahler, {\it Quantum Thermodynamics}, Springer (2004).
	\bibitem{QT2} E.~P.~Gyftopoulos and G.~P.~Beretta, {\it Thermodynamics: Foundations and Applications}, Dover (2005).
	\bibitem{QT3} A.~E.~Allahverdyan, R.~Ballian and Th.~M.~Nieuwenhuizen, J.~Mod.~Opt. {\bf 51}, 2703 (2004).
	\bibitem{W1} R.~Aliki, J.~Phys.~A {\bf 12}, L103 (1979).
	\bibitem{W2} R.~Kosloff and M.~A.~Ratner, J.~Chem.~Phys. {\bf 80}, 2352 (1984).
	\bibitem{W3} P.~Talkner, E.~Lutz and P.~H\"{a}nggi, Phys. Rev. E {\bf 75}, 050102(R) (2007).
	\bibitem{H1} F.~Tonner and G.~Mahler, Phys. Rev. E {\bf 72}, 066118 (2005).
	\bibitem{H2} M.~Henrich, M.~Michel and G.~Mahler, Europhys.~Lett. {\bf 76}, 1057 (2006).
	\bibitem{H3} M.~Henrich, G.~Mahler and M.~Michel, Phys. Rev. E {\bf 75}, 051118 (2007).
	\bibitem{H4} A.~E.~Allehverdyan and Th.~M.~Nieuwenhuizen, Phys.~Rev.~Lett. {\bf 85}, 1799 (2000).
	\bibitem{H5} T.~D.~Kieu, Phys. Rev. Lett. {\bf 93}, 140403 (2004).
	\bibitem{H6} T.~E.~Humphrey and H.~Linke, Physica E {\bf 29}, 390 (2005).
	\bibitem{H7} D.~Janzing {et. al.}, Int J.~Th.~Phys. {\bf 39}, 2717 (2000).
	\bibitem{C1} E.~Geva and R.~Kosloff, J. Chem. Phys. {\bf 96}, 3054 (1992); J. Chem. Phys. {\bf 104}, 7681 (1996).
	\bibitem{C2} T.~Feldmann and R.~Kosloff, Phys. Rev. E {\bf 61}, 4774 (2000).
	\bibitem{C3} J.~P.~Palao, R.~Kosloff and J.~M.~Gordon, Phys.~Rev.~E {\bf 64}, 056130 (2001).
	\bibitem{C4} D.~Segal and A.~Nitzan, Phys. Rev. E {\bf 73}, 026109 (2006).
	\bibitem{C5} G.~P.~Beretta, arXiv:quant-ph/0703.3261.
	\bibitem{C6} C.~M.~Bender, D.~C.~Brody and B.~K.~Meister, Proc.~R.~Soc.~Lond.~A {\bf 458}, 1519 (2002)
	\bibitem{T1} P.~Talkner, Annals of Phys. {\bf 167}, 390 (1986).
\end{thebibliography}
\end{document}